\documentclass[
    ,final            % use final for the camera ready runs
 ,numberedheadings % uncomment this option for numbered sections
  ]
  {aipproc}

\layoutstyle{8x11single}

\begin{document}

\title{Direct Reactions with Exotic Nuclei}

\classification{
%<Replace this text with PACS numbers; choose from this list:
%                \texttt{http://www.aip.org/pacs/index.html}
24.10.-i, 24.50.+g, 25.20.Lj, 25.60.-t}

\author{G.Baur}{
  address={Institut f\"{u}r Kernphysik, Forschungszentrum J\"{u}lich, 
    D-52425 J\"{u}lich, Germany}
}

\author{S.Typel}{
  address={Gesellschaft f\"{u}r Schwerionenforschung mbH (GSI),
           Planckstra\ss{}e 1, D-64291 Darmstadt, Germany}
}

\begin{abstract}
We discuss recent work on Coulomb dissociation
and an effective-range theory of 
low-lying electromagnetic strength of halo nuclei.
We propose to study Coulomb dissociation of a 
halo nucleus bound by a zero-range potential as a homework problem.
We study the transition from stripping to bound and 
unbound states and point out in this context that the 
Trojan-Horse method is a
suitable tool to investigate subthreshold resonances. 
\end{abstract}

\maketitle

%%%%%%%%%%%%%%%%%%%%%%%%%%%%%%%%%%%%%%%%%%%%
%% MAINMATTER
%%%%%%%%%%%%%%%%%%%%%%%%%%%%%%%%%%%%%%%%%%%%

\section{Introduction and Overview}

With the exotic beam facilities all over 
the world - and more are to come - direct reaction theories 
are experiencing a renaissance. 
We report on recent work - just finished 
and in progress - on 
Coulomb dissociation of halo nuclei \cite{tybaprl,tyba04}
and on transfer reactions to bound and scattering states.
We hope to report on further progress at the next 
workshop at MSU/ANL/INT/JINA/RIA or elsewhere.

Electromagnetic strength functions of halo nuclei exhibit 
universal features that can be described in terms of
characteristic scale parameters.
For a nucleus with nucleon+core structure
the reduced transition probability, as determined, e.g.,
by Coulomb dissociation experiments (for a review see
\cite{ppnp,hirsch}),
shows a typical shape that depends on the nucleon separation energy and the 
orbital momenta in the initial and final states.
The sensitivity to the final-state interaction (FSI) between the nucleon
and the core
can be studied systematically by varying the strength of the interaction
in the continuum. In the case of neutron+core nuclei  
analytical results for the reduced transition
probabilities are obtained by introducing 
an effective-range expansion.
The scaling with the relevant parameters is found explicitly.
General trends are observed by studying
several examples of neutron+core and proton+core nuclei
in a single-particle model assuming Woods-Saxon potentials.
Many important features of the neutron halo case can
be obtained already from a square-well model. Rather
simple analytical formulae are found.
The nucleon-core interaction in the continuum
affects the determination of astrophysical S factors at zero energy
in the method of asymptotic normalisation coefficients (ANC).
It is also relevant for the extrapolation of radiative capture
cross sections to low energies.

Coulomb dissociation of a neutron halo nucleus 
in the limit of a zero-range neutron-core interaction
in the Coulomb field of a target nucleus can be 
studied in various limits of the parameter space
and rather simple analytical solutions can be found. 
We propose to solve the scattering problem for 
this model Hamiltonian by means of 
the various advanced numerical methods that are available nowadays.
In this way their range of applicability can be studied
by comparison to the analytical benchmark solutions.

The Trojan-Horse Method \cite{tyba02,fus03} is a particular case
of transfer reactions to the continuum under quasi-free scattering
conditions.
Special attention is paid to the 
transition from reactions to bound and unbound states
and the role of subthreshold resonances.
Since the binding energies of nuclei close to the drip
line tend to be small, this is expected to be 
an important general feature in  
exotic nuclei. 

\section{Effective Range Theory of Halo Nuclei}
At low energies the effect of the nuclear potential is 
conveniently described by the effective-range expansion
\cite{Bet49}.
An effective-range approach for the electromagnetic
strength distribution in neutron halo 
nuclei was introduced in \cite{tybaprl} and
applied to the single neutron halo nucleus ${}^{11}$Be.
Recently, the same method was applied to the description of
electromagnetic dipole strength in ${}^{23}$O
\cite{Noc04}.
A systematic study sheds 
light on the sensitivity of the electromagnetic 
strength distribution to the interaction in the continuum.
We expose the dependence on the binding energy of the nucleon
and on the angular
momentum quantum numbers. Our approach extends the familiar textbook
case of the deuteron,
that can be considered as the
prime example of a halo nucleus, to arbitrary nucleon+core systems,
for related work see \cite{kala,besu,be}.
We also investigate in detail the square-well potential model.
It has  great merits: it can be solved analytically,
it shows the main characteristic features
and leads to rather simple and transparent formulae.
As far as we know, some of these formulae have not been published before.
These explicit results can be compared to our general
theory for low energies (effective-range approach) and also 
to more realistic Woods-Saxon models. Due to shape independence,
the results of these various approaches will not differ for
low energies. It will be interesting to delineate
the range of validity of the simple models.  

Our effective-range approach is closely related to effective field theories
that are nowadays used for the description of 
the nucleon-nucleon system and halo nuclei
\cite{Ber02}. 
The characteristic low-energy
parameters are linked to QCD in systematic expansions.
Similar methods are also used in 
the study of exotic atoms ($\pi^-A$, $\pi^+\pi^-$, $\pi^-p$, \dots) 
in terms of effective-range parameters.
The close relation of effective field theory to the effective-range
approach for hadronic atoms was discussed in Ref.\ \cite{Hol99}.

\begin{figure}
\includegraphics[height=.3\textheight]{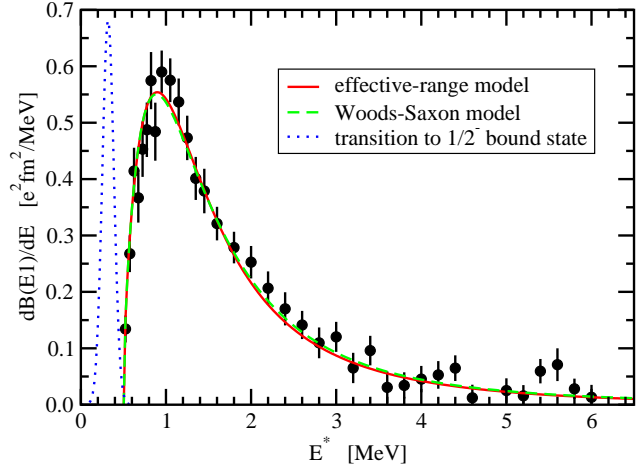}
\caption{\label{fig:1} Reduced probability for
dipole transitions as a function
of the excitation energy $E^{\ast}=E+S_{n}$
in comparison to experimental data extracted from
Coulomb dissociation of ${}^{11}$Be \cite{palit}.}
\end{figure}

In Fig.~\ref{fig:1} we show the application of the method to 
the electromagnetic dipole strength in $^{11}$Be. 
The reduced transition probability was deduced from high-energy
${}^{11}$Be Coulomb dissociation at GSI \cite{palit}.
Using a cutoff radius of $R=2.78$~fm and 
an inverse bound-state decay length of
$q=0.1486$~fm${}^{-1}$ as input parameters we extract
an ANC of $C_{0}=0.724(8)$~fm${}^{-1/2}$ 
from the fit to the experimental data. The ANC
can be converted to a spectroscopic factor of $C^{2}S=0.704(15)$
that is consistent with results from other methods.
In the lowest order of 
the effective-range expansion the phase shift 
$\delta_{l}^{j}$
in the partial wave with orbital angular momentum $l$ and
total angular momentum $j$
is written as $\tan \delta_{l}^{j}= -(x c_{l}^{j}\gamma)^{2l+1}$, 
where $\gamma=qR=0.4132<1$ is the halo expansion parameter and $x=k/q
=\sqrt{E/S_{n}}$ with the neutron separation energy $S_{n}$. 
The parameter  $c_{l}^{j}$ corresponds to the scattering
length $a_{l}^{j} = (c_{l}^{j}R)^{2l+1}$. We obtain
$c^{3/2}_{1}=-0.41(86,-20)$ and $c^{1/2}_{1}=2.77(13,-14)$. The latter 
is unnaturally large because of the existence of a bound $\frac{1}{2}^{-}$
state close to the neutron breakup threshold in ${}^{11}$Be.
For a further discussion we refer to \cite{tybaprl}.

\section{Homework problem}

We consider a three-body system consisting of 
a neutron $n$, a core $c$ and an (infinitely heavy) 
target nucleus with charge $Ze$. 
The Hamiltonian is given by
\begin{equation}
H=T+V_{cZ}+ V_{nc}
\end{equation}
where $T$ is the kinetic energy.
The Coulomb interaction between the core and the target
is given by $V_{cZ}=Z Z_{c} e^2/r_{c}$ 
and $V_{nc}$ is a zero-range interaction between 
$c$ and $n$. The s-wave bound state of the $a=(c+n)$ system
is given by the wave function $\Phi_0=\sqrt{q/(2\pi)}
\exp{(-qr)}/r$, where $q$ is related to the binding energy
$E_{b}$ by $ E_{b} = \hbar^{2}q^{2}/(2\mu)$ and
the reduced mass of the $c+n$ system
is denoted by $\mu$. We refer to \cite{ppnp} 
(see especially Ch.~4 there) for details. 
(The present homework problem is simpler than the one assigned
by I.~Thompson: in his case there is a p-wave
bound state in ${}^{8}$Li, and, in addition, the interactions between
the target and the projectile are much more complicated.)

One can study elastic scattering (influence of the polarisation
potential) as well as breakup of the halo nucleus $a$ 
in the Coulomb field of the target nucleus $Z$.
Although the Coulomb dissociation of this zero-range halo nucleus
is governed by a rather simple Hamiltonian,
the solution of this problem is nontrivial,
as is often the case in physics.
This model is also relevant for the Oppenheimer-Phillips 
process (polarisation of a deuteron in the Coulomb
field of a nucleus) \cite{opp}, see also \cite{ben} for a criticism. 
The parameter space is given by the charge $Ze$ of
the target and $Z_{c}e$ of the core $c$, 
the binding energy $E_{b}$ of the $(c+n)$ system,
the neutron and core masses $m_{n}$ and 
$m_{c}$ respectively. 

In this model one can study elastic scattering as well as 
breakup.  The beam momentum is denoted by $\vec{q}_{a}$ 
(the beam velocity is denoted by $v$), the 
momenta of the outgoing fragments $c$ and $n$
are $\vec{q}_{c}$ and $\vec{q}_{n}$, respectively (or $\vec{q}_{a}^{\prime}$
in the case of elastic scattering).
In the case of elastic scattering, the influence of the 
polarisation potential can be studied \cite{brt}.
The polarizability of a zero-range neutron halo nucleus
is given by 
\begin{equation}
 \alpha _{pol}=\frac{\hbar c}{2\pi^{2}}\sigma_{-2}
=\frac{(Z_c m_n e \hbar)^{2}}{6 \mu (m_n+m_c)^{2} E_{b}^{2}} \: .
\end{equation} 
For a small binding energy $E_{b}$ this can be a large effect.
In 1982 the electric dipole polarizability of the deuteron 
was determined by measuring elastic  scattering 
of deuterons on $^{208}$Pb at energies from 3.0 to 7.0 MeV
\cite{lynch}. (By the way, two of the authors of this paper were
participating in  this workshop.) 
The measured value of the electric polarizability
$\alpha_{pol}=(0.70 \pm 0.05)$~fm${}^{3}$ is in fair agreement with 
eq. 2, if the necessary finite range corrections are applied
see, e.g., \cite{wuo}.

The kinematics of the breakup process is given by 
$ \vec{q}_{a} \rightarrow \vec{q}_{cm} + \vec{q}_{rel}$ where
$\vec{q}_{cm}$ and $\vec{q}_{rel}$ are directly 
related to $\vec{q}_{c}$ and $\vec{q}_{n}$, respectively.
Analytic results are known for the plane-wave limit,
the Coulomb-wave Born approximation (CWBA,
``Bremsstrahlung integral'') and the adiabatic approximation
(Ron Johnson, this workshop and \cite{jeff}).
A first derivation  of the ``Bremsstrahlung formula'' was given by 
Landau and Lifshitz \cite{ll}, it was improved by 
Breit in \cite{breit}; an early review is given in \cite{bata}.

In the plane-wave limit the result does not depend on 
$q_{a}$ itself but only on the ``Coulomb push'' 
$\vec{q}_{coul}=\vec{q}_{a} - \vec{q}_{cm}$.

In the semiclassical high energy straight-line 
and electric dipole limit, first and second order
analytical results are available, as well as for the sudden limit.
E.g., in the straight-line dipole approximation 
a shape parameter $x=k/q$ and a strength parameter 
$y=m_{n}\eta/[(m_{n}+m_{c})b q]$ determine the 
breakup probability (in the sudden limit). The impact parameter
is denoted by $b$ and the Coulomb parameter is
$\eta=Z Z_c e^2/(\hbar v)$. 
In \cite{tyba01} it was found that the breakup probability
is given in leading order by 
\begin{equation}
\frac{dP_{LO}}{dk}=\frac{16}{3\pi q}
y^{2} \frac{x^{4}}{(1+x^{2})^{4}}
\end{equation}
and in next-to-leading order by 
\begin{equation}
\frac{dP_{NLO}}{dk}=\frac{16}{3\pi q}
 y^{4} \frac{x^{2}(5-55x^{2}+28x^{4})}{15(1+x^{2})^{6}}.
\end{equation}
Another important scaling parameter,
in addition to $x$ and $y$, is $\xi=\omega b/v$,
where $\hbar \omega$ is the excitation energy of the $(c+n)$ system.
In the sudden approximation we have $\xi=0$
and there is an analytical solution \cite{tybanuc}.

This homework problem can be studied, e.g., in the CDCC method,
which was widely discussed at the workshop.
It would be very interesting to see how well this method works
in various limits of the parameter space.
An especially interesting limit is the limit of low beam energies,
where the CWBA is very appropriate. We would expect that higher-order 
effects are very important under these conditions
and it would be good  to see that  the CDCC method converges.
We refer to \cite{ppnp}, especially Sect. 4.2, for 
further details and references on experimental and theoretical work
on  $E_{d}=12$~MeV deuteron breakup on ${}^{197}$Au. 

It would also be extremely interesting to apply the three-body
methods of \cite{aim} to the homework problem. In this work,
the so-called post-decay acceleration of the fragments is studied
and genuine three-particle wave functions for the final state
are used. In their case there are three charged particles
in the final state, but the problem is  non-trivial
even for only two (out of three) charged particles in the final state.

A related problem, the Coulomb breakup of antideuterons
bound in an orbit with quantum numbers $n$, $l$, $m$ 
\cite{eri,baur} can also be studied 
with this  Hamiltonian: in this case the charge of the core nucleus
$c$ is negative, $Z_{c}=-|Z_{c}|$. In \cite{eri}
the adiabatic method is used: the antideuteron c.m. motion
is assumed to be slow compared to the internal $\bar p$ and 
$\bar n$-motion and the authors calculate the antideuteron 
tunnel probability through the 
Coulomb barrier which is provided by the nucleus Z.

\section{Transfer Reactions}

Exotic nuclei have low thresholds for particle emission.
It is expected that in transfer reactions one will
often meet a situation where the transferred particle is 
in a state close to the particle threshold. 
In ``normal'' nuclei, the neutron threshold is
around an excitation energy of about 8 MeV, and 
the pure single particle picture is not directly applicable.
Much is known from stripping treactions like $(d,p)$
and thermal neutron scattering, see, e.g., \cite{bomo}.
The single particle strength
is fragmented over many more complicated compound states. The 
interesting quantity is the strength function 
which is proportional to  $\Gamma/D$
where $\Gamma$ is the width and D the level spacing.
This ratio is $\ll 1$, as can be estimated from a 
square well model (see, e.g., \cite{bomo}).
For $l=0$ there are no sharp
resonances, since $\Gamma>E$ around threshold. Due to the 
angular momentum (and/or) Coulomb barrier, one has
$\Gamma/E \ll 1$ at threshold for all the other cases.

For neutron rich (halo) nuclei the neutron threshold
is much lower, of the order of one MeV. In this
case the single-particle properties
are dominant and the ideas developed in the following can become relevant,
see also \cite{blan}. The level density is also much lower. 
In normal nuclei the level density at particle threshold is generally
so high that the single particle structure is very much
dissolved. This can be quite different in exotic nuclei
which can show a very pronounced single particle 
structure. 

\subsection{Trojan-Horse Method}

A similarity between cross sections for two-body and closely
related three-body reactions under certain kinematical conditions
\cite{Fuc71}
led to the introduction of the Trojan-Horse method 
\cite{Bau76,Bau84,Typ00,tyba02}.
In this indirect approach a two-body reaction
\begin{equation} \label{APreac}
 A + x \to C + c
\end{equation}
that is relevant to nuclear astrophysics is replaced by a reaction
\begin{equation} \label{THreac}
 A + a \to C + c + b
\end{equation}
with three particles in the final state. 
One assumes that the Trojan horse
$a$ is composed predominantly of clusters $x$ and $b$, i.e.\  $a=(x+b)$. 
This reaction can be considered as a special case of a transfer 
reaction to the continuum. It is studied experimentally under quasi-free
scattering conditions, i.e.\ when the momentum transfer to the
spectator $b$ is small. The method was primarily applied to the
extraction of the low-energy cross section of reaction
(\ref{APreac}) that is relevant for astrophysics. However, the method
can also be applied to the study of single-particle states in exotic
nuclei around the particle threshold. 

\subsection{Continuous Transition from Bound to Unbound State
Stripping}

Motivated by this 
we look again at the relation between transfer to 
bound and unbound states. Our notation is as follows:
we have the reaction
\begin{equation}
A+a \rightarrow B+b
\end{equation}
where $a=(b+x)$ and B denotes the final
$B=(A+x)$ system. It can be a bound state
$B$ with binding energy $E_{bind}=-E_{Ax}(>0)$,
the open channel $A+x, \mbox{with }E_{Ax}>0$, or 
another channel $C+c$ of the system $B=(A+x)$.
In particular, the reaction $x+A \rightarrow C+c$ can have
a positive $Q$ value and the energy $E_{Ax}$ can be negative
as well as positive.
As an example we quote the recently studied Trojan horse reaction
$d$+${}^{6}$Li \cite{auro03} applied to the ${}^{6}$Li$(p,\alpha)^{3}$He
two-body reaction (the neutron being the spectator).
In this case there are two charged
particles in the initial state (${}^{6}$Li+$p$).
Another example with a neutral particle $x$
would be ${}^{10}$Be$ + d \rightarrow p
+{}^{11}$Be$ +\gamma$.  
The general question which we want to answer
here is how the two regions $E_{Ax}>0$ and
$E_{Ax}<0$ are related to each other. 
E.g., in Fig. 7 of \cite{auro03} the coincidence yield 
is plotted as a function of the ${}^{6}$Li-$p$ relative energy.
It is nonzero at zero relative energy. How does 
the theory \cite{tyba02} (and the experiment)
continue to negative relative energies?
With this method, subtreshold resonances can be 
investigated rather directly.
We treat two cases separately, one where
system $B$ is always in the $(A+x)$ channel,
with a real potential $V_{Ax}$ between $A$ and $x$.
In the other case, there are also other channels
$C+c$, at positive and negative energies $E_{Ax}$. 

\subsubsection{One Channel Case}

We imagine the following situation:
The potential $V_{Ax}$ gives rise to a bound state 
with angular momentum $l$ close to threshold.
Now we decrease the potential so that the bound state disappears
and reappears as a resonance in the continuum. 
For $l>0$ there are sharp resonances and we can define a cross section
for stripping to a resonance by integrating over the resonance
line (over an energy range which is several times 
larger than the width) and they join 
smoothly to the stripping to the bound states, see \cite{bt74}.

Due to the absence of the angular momentum
barrier for $l=0$ there are some peculiarites which we study now.
Stripping to bound states is determined by the 
asymptotic normalization constant $B$ 
(see eqs. (A54) - (A56) of \cite{tyba04}) of the 
bound-state wave function and the function $h_{l}(iqr)$ where $q$
is related to the binding energy. Since
\begin{equation}
B \sim q^{3/2 }  \quad \mbox{for} \quad l=0
\end{equation}
and
\begin{equation}
B \sim q^{l+1}R^{l-1/2}  \quad \mbox{for} \quad l>0
\end{equation}
the stripping cross section (see, e.g., eq. (17) of \cite{bt74})
to a (halo) state with $l=0$ tends to zero for $q$ going to zero,
while it stays finite for $l>0$.
We note that the presence of a bound state close to zero 
energy leads to a large scattering length in the
$A+x$ system which leads to an enhancement of the 
elastic breakup cross section. 
The double differential cross section at threshold is 
proportional to
\begin{equation}
\frac{d^{2}\sigma}{d\Omega dE} \sim \frac{\sin^2\delta_0}{k} \: .
\end{equation}
The quantity $\sin^2\delta_0$ is given by $k^2/(q^2+k^2)$
for a bound and virtual state. Thus the double differential
cross section tends to zero like $k \sim \sqrt{E_n}$ for $l=0$.

When the strength of the potential is decreased, the 
bound state becomes a virtual state, which again 
leads to a very large scattering
length, see also \cite{blan}. 
In this context it seems interesting to note
that about 30 years ago a new type of threshold effects 
was predicted in \cite{bre73} (what is now called a halo state 
was referred to as a puffy state in those days).
Related to this is the qualitative difference of 
$l=0$ and $l>0$ in the location of the poles of the S matrix in the
complex plane \cite{nuzve,weidi}.
Only for $l>0$ there are
poles of the S matrix close to the real axis.

\subsubsection{Absorption at Zero Energy, Multichannel Case}

We follow the work of Ichimura, Austern, Vincent, and Kasano
\cite{iav,ki} who have studied the case $E_{Ax}>0$
and we now extend it to the case of $E_{Ax}<0$.
The exclusive case can be also studied by generalizing,
e.g., eq. (61) of \cite{tyba02}.
 
For positive energies $E_{Ax}$ the inclusive cross section
for $A+a \rightarrow b+X$ where $X$ is any state of the system $B=(A+x)$
consists of an elastic and inelastic component, see eq. (2.20) of
\cite{iav} or eq. (8) of \cite{ki}. For negative energies
$E_{Ax}$ the elastic breakup component is zero,
and only the inelastic component remains.
For positive energies this inclusive inelastic breakup
cross section is written as \cite{ki}
\begin{equation} \label{eq:inclu}
 \frac{d \sigma_{inel}}{d^{3}k_{b}}=\frac{(2\pi)^{4}}{\pi \hbar v_{a}}
 \int d^{3}r \: W(r)
 \left|\int G_x(\vec{r},\vec{r}^{\prime})
 \rho(\vec{r}^{\prime})d^{3}r^{\prime}\right|^{2} \: .
\end{equation}
The ``source term'' $\rho$ can be calculated from the 
distorted waves in the incident and final channel
and is given by eq. (3) of \cite{ki}. The Green's function in the
$x+A$ channel is given by $G_{x}$ and $W=-
%(!)
\mbox{Im} U(r)$
where $U$ is the optical potential (assumed to be local)
in the $x+A$ channel.

It is now our aim to give a meaning to $W$ and $G_{x}$ 
for negative energies $E_{Ax}$ and show that the cross
section behaves smoothly when going from positive energies
to negative energies. 

In \cite{ki} the Green's function is expanded in 
partial waves as
\begin{equation} \label{eq:green}
G_x(\vec{r},\vec{r}^{\prime})=-\frac{2m}{\hbar^2 k_x} 
\sum_{lm} \frac{f_l(r_<) h_l(r_>)}
{r r^{\prime}} Y_{lm}(\hat{r}) Y_{lm}^{\ast}(\hat{r}^{\prime})
\end{equation}
where $f_l$ and $h_l$ are regular and outgoing radial
wave functions in the potential $U$. 

The imaginary part $-W$ of the optical model potential 
is related to the partial wave reaction cross 
section $\sigma_l$ of $x+A$ scattering by
(this is eq. (26) of \cite{ki})
\begin{equation} \label{eq:reac}
 \int_0^{\infty}W(x)|f_l(r)|^2 dr=\frac{\hbar^{2} k_{x}}{2 m} \sigma_{l} \: .
\end{equation}
The total reaction cross section $\sigma_{reac}$ is 
given by $\sigma_{reac}=\sum_l (2l+1) \sigma_l$ and 
$\sigma_l$ is related to the imaginary part of the phase shift by
$\sigma_l= \pi[1-\exp(-4 \mbox{Im} \delta_l)]/k^{2}$.
We now derive this equation and generalize it to the case 
of negative energies $E_{Ax}$.
According to (A.20) in \cite{tyba04} we normalize the regular scattering 
wave function $g_l$ as (our normalization differs from the one of 
Ref. \cite{ki} by a factor of k)
\begin{equation}
g_l \rightarrow \frac{1}{2i}\left[\exp(2i\delta_{l}) u_{l}^{(+)}
 -u_{l}^{(-)}\right]
\end{equation}
valid for $r$ ouside the range of the potential.
The ingoing  and outgoing wave functions $u_{l}^{(\pm)}$
are given by
\begin{equation}
 u_{l}^{(\pm)}=x(-y_{l} \pm i j_{l})
\end{equation}
for neutrons and 
\begin{equation}
u_{l}^{(\pm)}=\exp(\mp i\sigma_l)(G_{l} \pm i F_{l})
\end{equation}
for charged particles, respectively.
The asymptotic behaviour is 
$u_{l}^{(\pm)} \rightarrow 
 \exp\left[\pm i\left(x-\eta ln(2x) -l\pi/2 \right)\right]$.
For positive energies $E_{Ax}>0$ we have $x=kr$.
By the usual procedure we obtain
\begin{equation}
 -2i\frac{2m}{\hbar^2}\int_0^{\infty}W(r) \left|g_{l}\right|^{2}dr
 =\left. \left( g_{l}^{\ast}\frac{dg_l}{dr} 
 - g_l \frac{dg_{l}^{\ast}}{dr}\right)\right|_{r=\infty} \: .
\end{equation} 
From the Wronskian relation $G \frac{dF}{dx}-F\frac{dG}{dx}=1$
we obtain $u_l^{(+)}\frac{du_l^{(-)}}{dx}-u_l^{(-)}\frac{du_l^{(+)}}{dx}=-2i$.
Using this we can evaluate the RHS. For positive 
energies we have $[u_l^{(\pm)}]^{\ast}=u_l^{(\mp)}$ and the RHS is given by
\begin{equation}
RHS=\frac{k}{2i}\left[1-\exp(-4\mbox{Im} \delta_l)\right] \: .
\end{equation}
This quantity is directly related to the partial 
wave reaction cross section $\sigma_l$ and eq. (\ref{eq:reac})
is established.
For low energies $E_{Ax}>0$ the phase shift is small and we can expand
\begin{equation}
 RHS=-2ik \mbox{Im} \delta_l \: .
\end{equation}
For negative energies $E_{Ax}<0$ we put $x=iqr$.
The functions $u_l^{(\pm)}$ are
exponentially decreasing and increasing respectively.
(A bound state corresponds to a pole of $S_{l}=\exp(2i\delta_l)$.) 
They are given asymptotically by (disregarding the logarithmic Coulomb phase)
\begin{equation}
u^{(\pm)}=i^{\pm l}\exp(\mp qr) \: .
\end{equation}
Using these properties we can evaluate the Wronskians
and the RHS is found to be
\begin{equation}
 RHS=\frac{q}{2}(-1)^l \left[\exp(2i\delta_l) -\exp(-2i\delta_l^{\ast})\right]
 \: .
\end{equation}
Close to the threshold $\delta_l$ is small and we have
\begin{equation}
 RHS=iq(-1)^l(\delta_l+\delta_l^{\ast})=2iq(-1)^l \mbox{Re} \delta_l \: .
\end{equation}

We can 
assume that 
the interior logarithmic derivative $L_i$ is smooth when
$E_{Ax}$ goes from positive to negative values. Now we can relate
the value of $\delta_l$ to this logarithmic derivative
and show in this way that the transition from positive to negative values
of $E_{Ax}$ is smooth. 
In the presence of an imaginary part $W$ the LHS is
non-vanishing. The logarithmic derivative $L_i$ is 
complex. This means that for $E_{Ax}>0$ 
$\delta_l$ acquires an imaginary part, for $E<0$ the
``phase shift'' $\delta_l$ acquires a real part.

Let us  deal with neutral particles. 
For low (positive) energies we can express the phase 
shift in terms of the scattering length $a_l$
by 
$ \tan(\delta_l) = -a_l k^{2l+1}$ where the 
scattering length is related to the interior logarithmic
derivative $L_i$ by eq.\ (A.31) of \cite{tyba04}
\begin{equation}
 a_l=a_{l}^{hs}\left(1-\frac{2l+1}{L_i+l}\right)
\end{equation}
where the hard sphere scattering length is given by
$a_l^{hs}=R^{2l+1}/[(2l+1)!!(2l-1)!!]$.
In order to obtain this result, the expansion of the 
Bessel  and Neumann functions for small values of 
$kr$ was used: $j_l=(kr)^l/(2l+1)!!$ and
$n_l=-(2l-1)!!/(kr)^{l+1}$.
We can write
\begin{equation} \label{eq:pos}
\delta_l=-k^{2l+1}a_{l}^{hs}\left(1-\frac{2l+1}{L_i+l}\right) \: .
\end{equation}
Thus the Wronskian can be expressed in terms of $L_i$.
For $E_{Ax}>0$ we find 
$RHS= -2i k^{2l+2}a_{l}^{hs}\mbox{Im} 
[(2l+1)/(L_i+l)]$.
For negative energies we put $k=-iq$ . Carrying through 
the corresponding steps as for the positive energy case
we obtain
\begin{equation} \label{eq:neg}
\delta_l=i(-1)^l q^{2l+1} a_{l}^{hs}
 \left(1-\frac{2l+1}{L_i+l}\right) \: .
\end{equation}
This leads to $RHS=-2iq^{2l+2} a_{l}^{hs} \mbox{Im}[(2l+1)/(L_{i}+l)]$. 
In our approach we have used the surface approximation, see eqs. (24)
and (25) of \cite{ki}. This means that the r-coordinate in eq. 
(\ref{eq:inclu}) is associated with the $r_<$-coordinate in eq.
(\ref{eq:green}) and $r^{\prime\prime}$ with $r_>$.
The $k^{2l+1}$ and $q^{2l+1}$ factors which enter in eqs.
(\ref{eq:pos}) and (\ref{eq:neg})
are cancelled by
the term coming from $h_l$, see eqs.
(\ref{eq:inclu}), (\ref{eq:green})
and eq. (25) of \cite{ki}. Thus there is a continuous
transition in the stripping from bound to unbound states.

Quite similarly, one can relate the logarithmic derivative
$L_i$ to the phase shift for charged particles
and establish the smooth transition from positive
to negative energies.
We do not give the details here.

\subsubsection{Imaginary part of the optical model potential
and solution of a toy model}

A formal expression for the optical potential is given 
in Eq. (2.16) of \cite{iav} by the Feshbach projector formalism.
In a schematic two-state model we want to illustrate
the smooth transition from positive to negative 
energies. We assume two channels with $l=0$,
the coupled radial equations are
\begin{equation} \label{eq:cc1}
\left(\frac{d^2}{dr^2}-u_1(r)+k_1^2\right)f_1(r)=u_{12}(r)f_2(r)
\end{equation}
and
\begin{equation} \label{eq:cc2}
\left(\frac{d^2}{dr^2}-u_2(r)+k_2^2\right)f_2(r)=u_{21}(r)f_1(r) \: .
\end{equation}
We have $k_2^2=k_1^2+Q(>0)$ and the channel 2 is open for 
$k_1^2=0 \mbox{ down to } k_1^2>-Q$. 
Introducing the Green's function $G_2(r,r^\prime)$ we can 
express $f_2$ as $f_2(r)=\int G_2(r,r^\prime)u_{21}
f_1(r^\prime)dr^\prime$.
Inserting this into eq. (\ref{eq:cc1}) we obtain 
an equation for $f_1$ in an optical potential.
This optical potential has a real and an imaginary part.
We are 
especially interested here in the imaginary part which 
can be found as follows: 
We can express the Green's function as 
$G_2=\int dE \: \chi_E(r) \chi_E(r^\prime)/(E^{+}-E)$.
Using $\lim \frac{1}{x-x_0\pm i\epsilon}=PP\frac{1}{x-x_0}
\mp i \pi \delta(x-x_0)$ we obtain $\mbox{Im} G_2=-i\pi \chi_E(r)
\chi_E(r^{\prime})$ where $\chi_E(r)$ is the regular solution
of the homogeneous part of eq. (\ref{eq:cc1}) 
(with the coupling potential $u_{21}=0$).
This leads to a nonlocal, separable imaginary part 
given by $W(r,r^\prime)=-\pi V_{12}(r)\chi_E(r) \chi_E(r^\prime)
V_{21}(r^\prime)$.

It is instructive to solve eqs. (\ref{eq:cc1}) and 
(\ref{eq:cc2}) analytically for a square-well model with 
delta-function coupling.
We take $u_1=-|u_1|, u_2=-|u_2|$ 
for $r<R$ and zero otherwise and $u_{12}=u_{21}
=u\delta(r-R)$. This leads to a Sprungbedingung in the
logarithmic derivatives. According to eqs. (22) ff. of \cite{tyba02}
we have the following asymptotic behaviour of 
the (s-wave) radial wave functions:
\begin{equation}
f_1(r) \rightarrow \frac{i}{2}\left[S_{12}^{\ast}\exp(-ik_1r)\right]
\end{equation}
and 
\begin{equation}
f_2 \rightarrow \frac{i}{2} \sqrt{\frac{v_2}{v_1}}
 \left[S_{22}^{\ast}\exp(-ik_2r)-\exp(ik_2r)\right] \: .
\end{equation}
The two logarithmic matching conditions determine
$z_1= \sqrt{k_2/k_1}S_{12}^{\ast}$ and $z_2=S_{22}^{\ast}$.
The interior logarithmic derivatives $L_1$ and $L_2$
are real (somewhat differently from the 
previous subsection they are defined here as
$L_{i}=f_{i}^{\prime}/f_{i}, i=1,2$). 
Introducing $\tilde{L}=L_2-u^2/(L_1+ik_1)$
one can express
$z_2=\exp(2ik_2R)(\tilde{L} -ik_2)/(\tilde{L}+ik_2)$
and $z_1= 2ik_2u \exp[i(k_1+k_2)R]/[(L_1+ik_1)(\tilde{L}+ik_2)]$.
From these expressions one can derive the unitarity of the S matrix
(2 by 2 for $k_1^2>0$).
The S-matrix element ($k_1^2>0$) $S_{12}$ has the threshold behaviour
$S_{12} \propto \sqrt{k_1}$ which is characteristic for the s wave.
It should be straightforward to generalize to $l>0$ and to
Coulomb interactions.

For $k_1^2<0$ there is only one 
open channel (channel 2) and the S matrix 
consists only of one S-matrix element $S_{22}$.
We put $k_1=-iq$ ($|E_n|=\hbar^2 q^2/(2m)$).
One sees that $\tilde{L}$ is real (rather than complex for the 2 channel case)
and $z_2$ is unitary (modulus is one). The quantity $z_1$
tends to a well defined number, of interest for the THM method.
For $E_n=0$ it is given by $z_1=2ik_2u\exp(ik_2R)/[L_1(\tilde{L}+ik_2)]$.
Since channel 1 is closed, $S_{12}$
is not an S-matrix element, but it can still be used as an input
in eqs. (64), (65) of \cite{tyba02}. The quantitiy $J_{l}^{(+)}$ there can
also be defined for imaginary values of $k_{Ax}$ (closed channel
case). 

\section{Conclusion}

While the foundations of direct reaction theory
have been laid several decades ago, the new possibilites which have 
opened up with the rare isotope beams 
are an invitation to revisit this field. The general frame 
is set by nonrelativistic many-body quantum scattering theory,
however, the increasing level of precision demands
a good understanding of relativistic effects notably in 
intermediate energy Coulomb excitation, see the talk by Carlos 
Bertulani at this workshop.

The properties of halo nuclei 
depend very sensitively on the binding energy 
and despite  the ever increasing
precision of microscopic approaches using
realistic NN forces it will not be possible, say,
to predict the binding energies of nuclei to a level of 
about 100 keV.
Thus halo nuclei ask for 
new approaches in terms of some effective 
low-energy constants.
Such a treatment was provided in Ch.\ 2 and an example
to the one-neutron halo nucleus $^{11}$Be was given.
With RIA one will be able to study also neutron halo
nuclei for intermediate mass nuclei. This is expected to
be relevant also for the astrophysical r process.
It is a great challenge to extend the present approach
for one-nucleon halo nuclei to more complicated cases,
like two-neutron halo nuclei.

The treatment of the 
continuum is a general problem,
which becomes more and more urgent when the dripline is approached.
In the present proceedings we studied 
the transition from bound to unbound states
as a typical example.

\begin{theacknowledgments}
We wish to thank Carlos Bertulani, Kai Hencken, Radhey Shyam
and Dirk Trautmann for their collaboration on various topics 
in this field.   
\end{theacknowledgments}

%%%%%%%%%%%%%%%%%%%%%%%%%%%%%%%%%%%%%%%%%%%%%%%%
%% The bibliography can be prepared using the BibTeX program or
%% manually.
%%
%% The code below assumes that BibTeX is used.  If the bibliography is
%% produced without BibTeX comment out the following lines and see the
%% aipguide.pdf for further information.
%%
%% For your convenience a manually coded example is appended
%% after the \end{document}
%%%%%%%%%%%%%%%%%%%%%%%%%%%%%%%%%%%%%%%%%%%%%%%%

%%%%%%%%%%%%%%%%%%%%%%%%%%%%%%%%%%%%%%%%%%%%%%%%
%% You may have to change the BibTeX style below, depending on your
%% setup or preferences.
%%
%%
%% For The AIP proceedings layouts use either
%%%%%%%%%%%%%%%%%%%%%%%%%%%%%%%%%%%%%%%%%%%%

\bibliographystyle{aipproc}   % if natbib is available
%\bibliographystyle{aipprocl} % if natbib is missing

%%%%%%%%%%%%%%%%%%%%%%%%%%%%%%%%%%%%%%%%%%%
%% You probably want to use your own bibtex database here
%%%%%%%%%%%%%%%%%%%%%%%%%%%%%%%%%%%%%%%%%%%
%\bibliography{sample}

%%%%%%%%%%%%%%%%%%%%%%%%%%%%%%%%%%%%%%%%%%%
%% Just a reminder that you may have to run bibtex
%% All of it up to \end{document} can be removed
%% if you don't like the warning.
%%%%%%%%%%%%%%%%%%%%%%%%%%%%%%%%%%%%%%%%%%%
%\IfFileExists{\jobname.bbl}{}
% {\typeout{}
%  \typeout{******************************************}
%  \typeout{** Please run "bibtex \jobname" to optain}
%  \typeout{** the bibliography and then re-run LaTeX}
%  \typeout{** twice to fix the references!}
%  \typeout{******************************************}
%  \typeout{}
% }
%
%%%%%%%%%%%%%%%%%%%%%%%%%%%%%%%%%%%%%%%%%%%
%% The following lines show an example how to produce a bibliography
%% without the help of the BibTeX program. This could be used instead
%% of the above.
%%%%%%%%%%%%%%%%%%%%%%%%%%%%%%%%%%%%%%%%%%%

\end{document}